\documentclass[]{spie}  
\usepackage[dvips]{graphicx} 

\title{The EXPLORE Project: \\
A Deep Search for Transiting Extra-Solar Planets \footnote{ 
\quad Based on observations collected from: the Cerro Tololo Inter-American 
Observatory, National
Optical Astronomy Observatories, which are operated by the Association of Universities
for Research in Astronomy, under contract with the National Science Foundation;
the European Southern Observatory, Chile, ESO proposal N$^\circ$ 267.C-5723;
the Canada-France-Hawaii Telescope, which is operated by
the National Research Council of Canada, Le Centre National de Recherche Scientifique
of France, and the University of Hawaii USA;  and
the W.M. Keck Observatory, which is operated as a
scientific partnership among the California Institute of Technology,
the University of California and the National Aeronautics and Space
Administration, and which was made possible by the generous
financial support of the W.M. Keck Foundation.}
}

\author{H.K.C.~Yee\supit{a}, G.~Mall\'en-Ornelas\supit{b}, 
S.~Seager\supit{c}, M.~Gladders\supit{d}, T.~Brown\supit{e}, \\
D.~Minniti\supit{f}, S.~Ellison\supit{g}, and
G.M.~Mall\'en-Fullerton\supit{h}
\skiplinehalf
\supit{a}University of Toronto, Toronto, Canada \\
\supit{b}Princeton University, Princeton, NJ; Pontificia Univ.~Cat\'olica, Santiago,
 Chile \\
\supit{c} Institute for Advanced Study, Princeton, NJ \\
\supit{d} Carnegie Obs., Pasadena, CA; Univ.~of Toronto, Toronto, Canada \\
\supit{e} High Altitude Observatory, NCAR, Boulder, CO\\
\supit{f} Pontificia Univ.~Cat\'olica, Santiago, Chile \\
\supit{g} European Southern Observatory, Santiago, Chile\\
\supit{h} Universidad Iberoamericana, Mexico City, M\'exico
}

\authorinfo{ Further author information:
Contact: H.K.C.Y:  E-mail: hyee@astro.utoronto.ca.~~
Current address: S.S.: DTM, Carnegie Institute of Washington, Washington, DC}

\begin{document} 
  \maketitle 

\begin{abstract}
  Searching for transits provides a very promising technique for 
finding close-in extra-solar planets. 
Transiting planets present the advantage of allowing one
to determine physical properties such as mass and radius unambiguously.  
The EXPLORE (EXtra-solar PLanet Occultation
REsearch) project is a transit search project carried out using
wide-field CCD imaging cameras on 4-m class telescopes, and 8--10m
class telescopes for radial velocity verification of the photometric candidates.
We describe some of the considerations that go into the design of
the EXPLORE transit search to maximize the discovery rate and minimize
contaminating objects that mimic transiting planets.
We show that high precision photometry (2 to 10 millimag) 
and high time sampling (few minutes) are crucial
for sifting out contaminating signatures, such as grazing binaries. 
We have an efficient data reduction pipeline  which allows us to 
completely reduce the data and search for transit candidates in less than 
one month after the imaging observations, allowing us to conduct
same-semester radial velocity follow-up observations, reducing the phase 
uncertainty.

We have completed two searches using the 8k MOSAIC camera at the 
CTIO4m and the CFH12k camera at CFHT, with runs covering 11 and 
16 nights, respectively. Using the 4400 images from the two fields, we 
obtained preliminary light curves for approximately 47,000 stars with better 
than $\sim1$\% photometric precision.  A number of light curves with
flat-bottomed eclipses consistent with 
being produced by transiting planets has been discovered.  
Preliminary results from 
follow-up spectroscopic observations using  the VLT UVES spectrograph and
the Keck HIRES spectrograph obtained for a number of the candidates are
presented.
Data from four of these can be interpreted consistently as possible 
planet candidates, although further data are still required for definitive
confirmations.

\end{abstract}


\keywords{Extra-solar planets, transits, photometry, radial velocity}

\section{INTRODUCTION}
\label{sect:intro}  

The discovery of giant extra-solar planets,
such as 51 Peg b \cite{mayor95}, with orbital
periods of a few days and orbital radius $<0.1$AU, was completely
unexpected.
Currently, 13 of these close-in extra-solar giant planets (CEGPs)
are known, representing about 15\% of the planets discovered
by the radial velocity (RV) technique.\footnote{ See, e.g., 
Extrasolar Planets Encyclopaedia, http://www.obspm.fr/encycl/catalog.html.}
The existence of close-in giant planets shows that planetary systems
can be radically different from our own, and sparked much
theoretical work on planet formation and migration scenarios to
explain the proximity of giant planets to the parent stars (e.g,
Refs. \citenum{lin96}, \citenum{holman97}, \citenum{murray98},
and \citenum{rasio96}).
This new class of planets
also makes the method of finding planets via the transiting of their
parent star very promising. 
In the situation of a Jupiter-sized planet transiting a solar-sized
parent star, an eclipse with a flat-bottom light curve
(due to the planet being completely superimposed on the parent
star) of depth $\sim$1\% is expected, an easily measurable
effect with modern CCD photometry.
The probability that a given planet
will show transits is inversely proportional to its orbital distance;
for CEGPs, this is typically $\sim10$\%.
Moreover, the typical periods of a few days for CEGPs also makes
monitoring for transits relatively easy.

Transiting planets offer a number of advantages over those discovered
by the RV technique alone.
They are currently the only ones for which a radius can be measured.
Furthermore, absolute masses of transiting planets can be measured
using the RV technique without  the usual sin $i$ ambiguity.
Transiting planets are also the most suitable for many kinds of
follow-up studies, e.g., atmosphere transmission 
spectroscopy\cite{seager00}\cite{char02},
searches for moons and rings\cite{brown01}, and others.

Over the past few years, a large number of transit searches
for extra-solar planets have been started (e.g., Refs. \citenum{vulcan},
\citenum{stare}, \citenum{ogle}).
However, currently only one unambiguous transiting planet is known,
HD 209458b, which was discovered originally by the RV technique and
later  found to be a transiting planet\cite{char00}\cite{henry00}.
In this paper we present preliminary results from the EXPLORE
(EXtra-solar PLanet Occultation REsearch) project, which is the
first 4m-class telescope transit search with 8-10m class
spectroscopic radial velocity follow-up observations
as an integral component of the search strategy.
In Section 2 we outline some of the considerations that went into
the design of the searches.
Section 3 describes the two EXPLORE searches conducted so far.
Preliminary results and possible transiting planet candidates 
are presented in Section 4.
Finally, in Section 5
we provide a summary and briefly outline the future prospects
 for the EXPLORE project.

\section{Some Considerations for the Design of Transit Surveys} 

Detailed examinations of the various considerations that went into
the design of the EXPLORE searches can be found in Refs. \citenum{exp1}
and \citenum{unique02}.
Here, we present a brief summary.

\subsection{Planet Detection Probability}

An estimate of the fraction of monitored stars that have short-period
transiting planets is instructive for designing a transit survey.
The fraction of stars with transiting planets can be approximated
by $F_p\sim P_pP_g$, where $P_p$ is the probability that a star has
a planet within certain orbital radius, and $P_g$ is the geometric
probability that the planet will produce an eclipse.
Using $P_p\sim0.007$ (Ref. \citenum{butler01}) and $P_g\sim0.1$ for
close-in planets of approximately Jupiter size, and assuming we can 
detect planets only around
isolated stars (adopting a binary fraction of 1/2), we get $F_p\sim
0.0035$, or one transiting CEGP in $\sim$3000 stars monitored with
photometric precision of better than $\sim$1\%.

In practice, the fraction of stars with detected planetary transits
will be much smaller (by a factor of 2 to 10) in a transit search, 
and it is dependent on a
number of observational parameters, such as the window function
of the observations, the photometric precision, and the time sampling.
We discuss the effects of some of these parameters in the subsections below.

\subsection{Constraining Planet Candidate Properties}

An attractive aspect of transit searches is that with a high precision
and high time-sampling light curve 
showing two or more eclipses with flat bottoms,
a unique solution of the orbital parameters and companion radius
can be derived by assuming a circular orbit and a mass-to-radius
relation for the parent star.
Specifically, the light curve needs to be of sufficient quality so
that the time of ingress and egress, and the duration of the flat-bottomed
part can be measured.
The details concerning the derivations of the various parameters 
(specifically: the mass of the star, the radius of the star 
and of the companion, orbital distance, and orbit inclination) based
on such a light curve can be found in Refs. \citenum{exp1}~and
\citenum{unique02}.
The unique solution provides a powerful tool to eliminate contaminating
candidates (see Section 2.3).
Thus, high-quality light curves are crucial for selecting the most
robust planet candidates for mass determination follow-up observations,
significantly increasing the efficiency of large telescope 
spectroscopy time.

\subsection{Ruling out Contaminating Systems}
The key signatures for a planet transit are:
(1) they show a shallow eclipse (a few percent at the most),
(2) the eclipses have a flat bottom (in a color where limb darkening 
is negligible), and (3) there is no secondary eclipse.
However, there are at least three types of systems which can be 
confused with a transiting planetary system. Fortunately,
high precision, high time-sampling light curves often can be used to
rule out these contaminants, allowing one to produce a highly robust
candidate list which will provide a high yield in radial velocity
follow-up confirmation.  We discuss these three cases briefly below.
Examples of
actual cases of these contaminants from our searches are presented
in Section 4.

\noindent
(1) {\it Grazing Eclipsing Binaries}:
At certain orbital inclinations, an eclipsing binary star system can produce
the typical drop of $\sim$1\%~expected for a transiting planet.
For systems with two similar temperature stars, it would be impossible
to discern a secondary eclipse.
However, given a sufficiently high signal-to-noise ratio
and high time-sample light curve, 
the shape of a grazing binary system can be differentiated
from a planet transit, as the former would have V- or U-shape
eclipses, whereas the latter would have a flat-bottomed eclipse.
Distinguishing the two cases is also the easiest when limb-darkening
is minimized, as in the case for light curves obtained in the redder
pass bands.

\noindent
(2) {\it Eclipsing Binary Systems with a Large Primary Star}:
A small star eclipsing a much larger star can produce a flat-bottomed
light curve with $\sim1$\%~depth, mimicking a typical expected
signal from a transiting planet around an average-sized star.
Light curves with sufficient information to constrain a
unique solution to the 5 parameters listed in Section 2.2
are usually able
to distinguish such contaminants.
For example, such a system often has a relatively
long transit time.
Spectral typing of the primary star will also allow one to 
differentiate such a system from a bona fide planetary transit.

\noindent
(3) {\it Presence of a Contaminating Blended Star}:
A flat-bottomed and relatively deep eclipse from a companion star
fully superimposed on its larger primary will appear shallower
if there is a third, brighter star blended in the light curve.
This type of contamination is probably the most difficult of
the three to rule out based solely on the light curve.
Typically such a blended system will have a relatively small
ratio of flat-bottom to ingress/egress time, giving the
appearance of a planet transiting within the narrow range of
impact parameter near the stellar limb, which
statistically is a relatively rare occurrence.
In this situation, the 
solution to the system derived from the light curve
may not be able to rule out such a contaminant definitively,
and an additional constraint in the form of the radius of the
primary star from spectral classification would be needed.
If radial velocity data are available, such a system will
typically appear as having a dominant zero velocity component
(from the bright blended star) with a weaker component from
the primary of the eclipsing system with the period and phase
corresponding to the observed eclipses.

We note that a fourth, and potentially large, contaminating class is 
eclipsing late-M and brown dwarfs which have radii similar to gas giant
planets.  Such contaminants can only be ruled out by mass estimates
using spectroscopy (see Section 2.5).
However, most of these objects would be of interest in their own right.

\subsection{Maximizing the Visibility Probability}

To effectively detect a transit candidate, especially with sufficient
information to derive constraints on the characteristics of the system
(viz Section 2.2), we need to detect at least two full eclipses from a system.
For ground-based observations, using non-dedicated telescopes (e.g.,
4m-class telescopes) with a limited period of time assigned, it is
important to design searches which maximize the probability (which
we term the visibility probability, $P_{vis}$) that the two-eclipse
requirement is met.
From general considerations and simulations,
(for details, see Ref. \citenum{exp1}),
we find that using a continuous block of telescope time and following
the monitored field for as long as possible each night provide the
most efficient strategy.
Figure 1a illustrates examples of $P_{vis}$ of detecting transits with
different orbital periods for runs of different total lengths.
The total efficiency of a run can be quantified approximately by 
$P_{vis}$ integrated over the periods of interest.
For example, with $10.8$ hrs of monitoring per night, the integrated
$P_{vis}$ for systems
with periods between 2 to 5 nights, ranges from $\sim0.2$ for an 11-night
run to $\sim0.6$ for an 18-night run (Figure 1b).
In terms of efficiency per night ($<P_{vis}>$ per night), the optimal
efficiency peaks at about 20 nights, with a broad distribution between
16 to 30 nights, as shown in Figure 1c.

   \begin{figure}
   \begin{center}
   \begin{tabular}{c}
   \includegraphics[height=9cm]{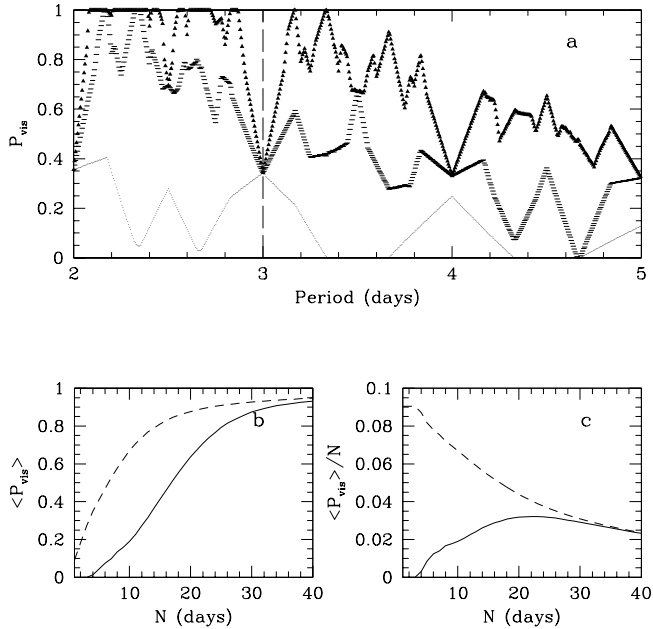}
   \end{tabular}
   \end{center}
   \caption[pvis]
   { \label{fig:pvis}
Panel a:
Examples of the visibility probability $P_{vis}$ of 
detecting transiting planets with
different orbital periods.
$P_{vis}$ is calculated with the requirement that two transits must be
observed.
Consecutive nights of 10.8 hours monitoring is assumed, with the triangles
representing a 21-night run; bars, a 14-night run; and dotted line, a 7-night
run, which is similar to the actual time coverage of the EXPLORE I search
after accounting for weather.
Panel b shows the integrated $P_{vis}$ as a function of the length of
the run based on continuous monitoring of 10.8 hrs per night, for the cases
of detecting two full eclipses (solid line) and one eclipse (dashed line). 
Panel c illustrates the efficiency per night, expressed in integrated
$P_{vis}$ per night, as a function of the number of nights.
}
   \end{figure}

\subsection{Radial Velocity Follow-up}

Late-M dwarfs ($M\ge80M_J$), brown dwarfs ($13M_J<M<80M_J$), and
gaseous giant planets ($M\le13M_J$) all have similar sizes due
to a competition between Coulomb force and electron
degeneracy effects\cite{hubbard02}. 
Hence, a radius measurement from the transit light curve alone is 
insufficient for determining whether the transiting body is a
true planet, even if it has a radius $\sim R_J$.
Radial velocity measurements to estimate the mass are therefore 
required to definitively determine the nature of the eclipsing object
in most cases.
Mass limits for
candidates from deep transit searches using 4m-class telescopes can
be routinely obtained using an echelle spectrograph on an 8m-class
telescope for relatively faint stars ($I\le18$).
For example, a solar mass star with an 80$M_J$ M dwarf or a 13 $M_J$
brown dwarf companion with an orbital radius of 0.05 AU will show RV
amplitudes of 10.1 km s$^{-1}$ and 1.6 km s$^{-1}$, respectively,
which can be ruled out easily with RV measurements with
 accuracies of several hundred meters per second.
Furthermore, if the period and phase of the orbit are known accurately, 
only a handful of RV points judiciously sampled are required to 
constrain the companion's mass (see Ref. \citenum{exp1}).


\section{The EXPLORE Project} 
\label{sect:explore}

Two EXPLORE searches were conducted in 2001, one using the CTIO 4m
telescope, and the other, the CFHT 3.6m.
The combination of
the wide-field capability of the mosaic CCD camera ($\sim$ 1/3 square
degrees) and the large
telescope aperture makes observing high star density fields on the
Galactic Plane an optimal strategy for maximizing the number of stars
monitored.
We use $I$ band images which minimize Galactic extinction and increase
the number of late-type (small) stars monitored.
Moreover, a red band decreases the effect of limb-darkening on the 
light curve shape of the eclipse, making a flat-bottomed eclipse more
recognizable.

\subsection{The EXPLORE I Search}

The EXPLORE I search was carried out at the CTIO 4m, 30/May and 
1--10/June/2001, for 11 nights.
We used the
Mosaic II CCD Imager, a 8096$\times$8096 pixel CCD camera with a pixel
scale of 0.27$''$/pixel, providing a field size of 36$'$ squared.
We observed a single field near the Galactic Plane with 
$l=-27.8^{\rm \circ}$ and $b=-2.7^{\rm \circ}$.
Images in $I$ band with a nominal integration time of 60 seconds
were taken  continuously.
With a readout time of 102 seconds, the sampling rate is 2.7 minutes.
The integration time was adjusted up or down based on seeing, cloud
conditions, and sky brightness.
We had clear weather for approximately 6 nights, with some additional
data under poor conditions.
Because of the loss of a significant number of nights to poor weather,
the integrated $P_{vis}$ for the run is very low, only $\sim7$\%.
Figure 2 is a gray-scale image of a 60$''$ section of the field,
showing the typical crowdedness of the field.

   \begin{figure}
   \begin{center}
   \begin{tabular}{c}
   \includegraphics[height=8cm]{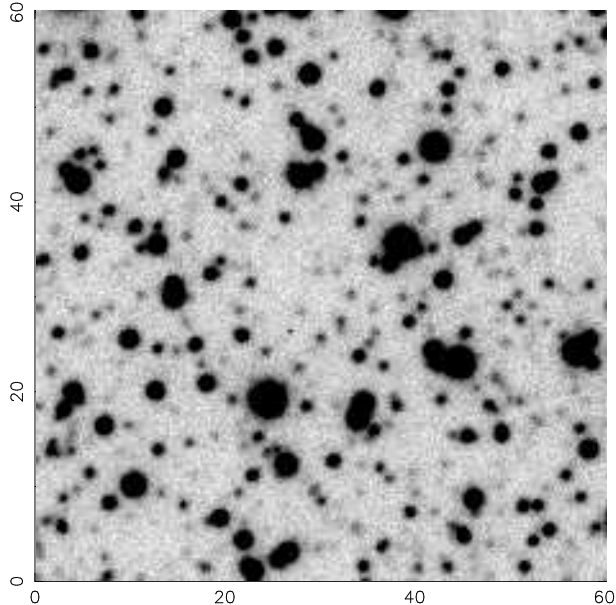}
   \end{tabular}
   \end{center}
   \caption[grayscale]
   { \label{fig:grayscale}
A gray scale image of a small section of the EXPLORE I field.
Each side is 60$''$$\times$60$''$.
The complete field of $\sim$36$'$ squared contains approximately
400,000 detected objects with $\sim100,000$ stars brighter than $I=18.2$.
}
   \end{figure}

A total of 1800 science frames were obtained from the EXPLORE I run.
We developed a customized pipeline optimized for performing high-precision
aperture photometry of faint stars in dense fields with well sampled 
point-spread functions.
Since EXPLORE monitors relatively faint stars (comparable to sky brightness), 
a key feature in the photometry algorithm is to use small 
apertures (2$''$ to 3$''$) to minimize sky noise.
In order to obtain accurate relative photometry using a small aperture, 
we use an iterative
sinc-shift algorithm to resample every star in such a way that
identical aperture, relative to the star profile, is used for integrating
the flux.
This is equivalent to centering the aperture on each star extremely
accurately, a crucial step in accurate relative photometry.
The data pipeline, from pre-processing of the images to the final 
production of the light curves, will be described in detail in Yee et 
al.~(in preparation).
Typically, our highly automated reduction procedures and pipelines
allow us to produce preliminary light curves within one to two weeks
of the end of an observing run, and a sample of possible transit 
candidates for spectroscopic follow-up within four weeks.

For the EXPLORE I field, we examined preliminary light curves of 
$\sim$37,000 stars with
photometry accuracy of better than 1\%.
Examples of light curves from the EXPLORE I search are shown in Figure 3
and discussed in Section 4.
The planet candidates were searched using visual inspection, in which
light curves with two or more shallow eclipses ($<3\%$) with discernible
flat bottoms are chosen as candidates.
While this is a relatively time consuming method, it is found,
 using a large number of simulated light curves which match the 
noise and time-sampling characteristics of the data, that visual inspection is
able to pick out planetary transits at the 100\% completeness level
when the the depth of the eclipse is 2.5 to 3 times the rms scatter
of the photometry of the light curve\cite{blee02}.

   \begin{figure}
   \begin{center}
   \begin{tabular}{c}
   \includegraphics[height=9cm]{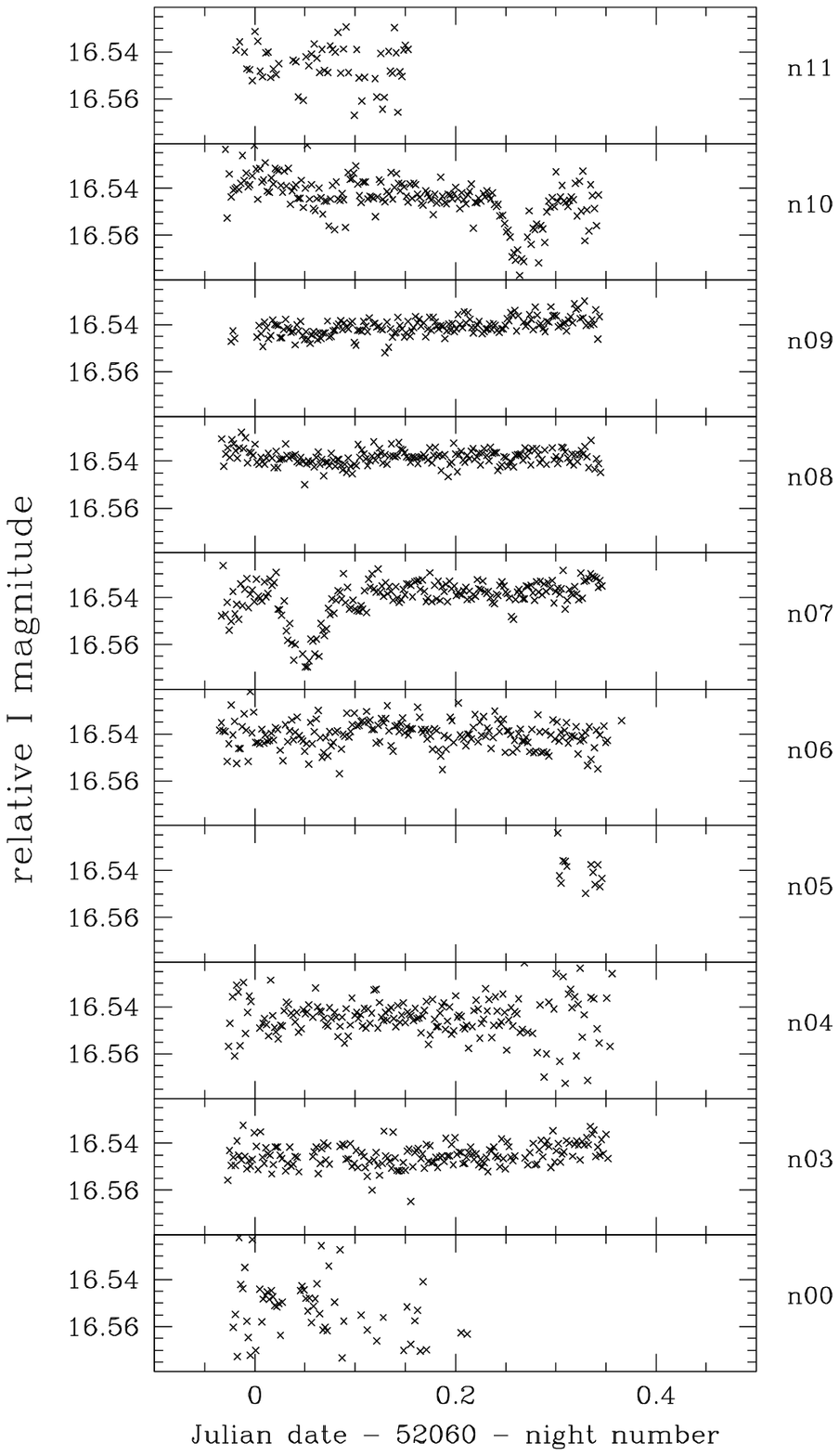}
   \includegraphics[height=9cm]{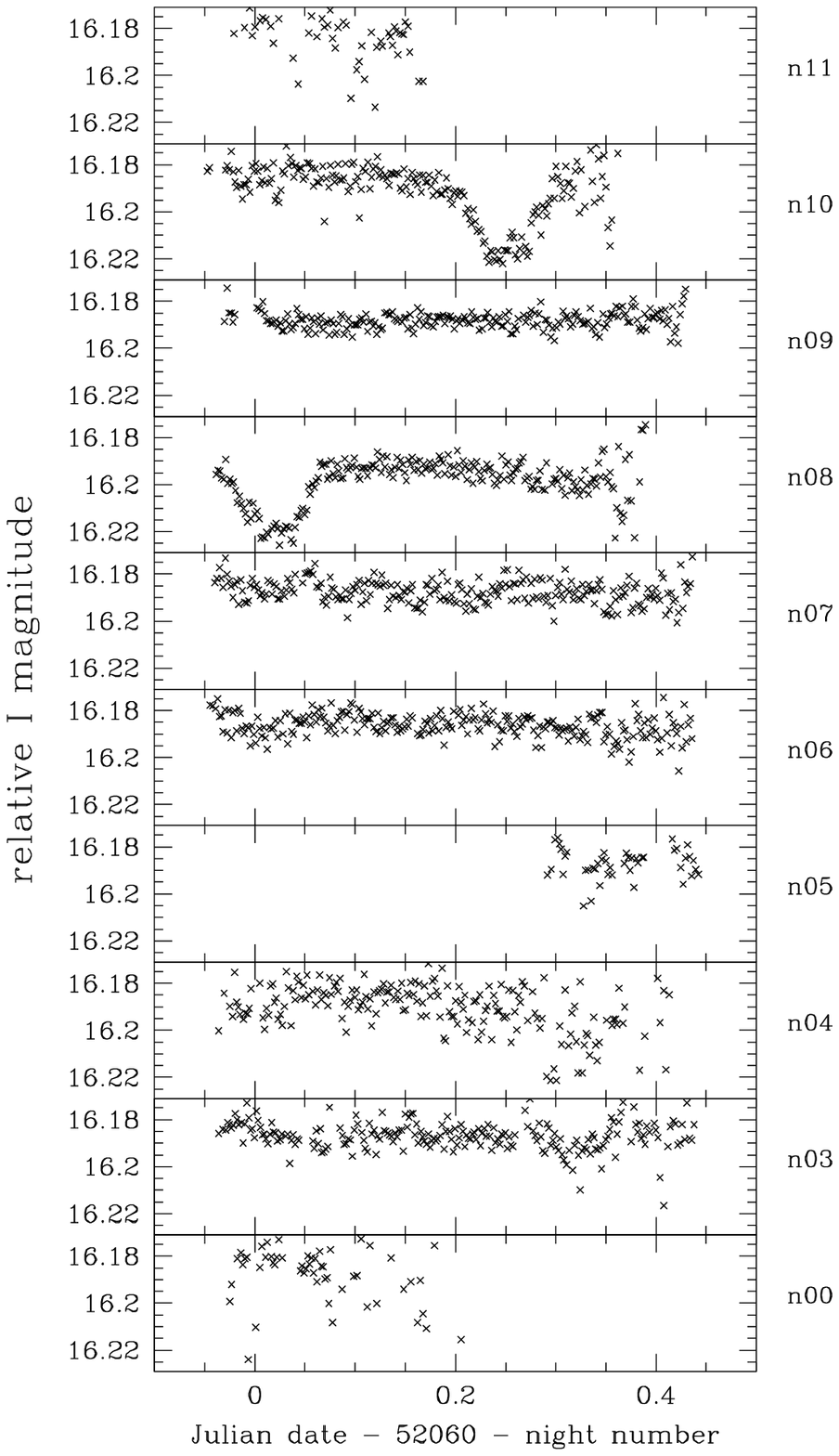}
   \end{tabular}
   \end{center}
   \caption[grazing]
   { \label{fig:grazing}
Example light curves from the EXPLORE I search.
Each box in the panels represents one night of monitoring, with time
expressed in relative Julian date.  The night number is marked on
the right hand side of the panels.
Left panel: EXP1-c07s5157:
The two 3\% eclipses are likely due to a grazing binary system, as
they have the typical V-shape expected for such systems.
Right panel: EXP1-c01s52805:
This object shows two eclipses of $\sim3$\% depth with clear flat
bottoms. However, the relatively small ratio of the flat-bottom
to ingress/egress time of indicates that the object is likely
a blend of an eclipsing system plus a third brighter star.
}
   \end{figure}

\subsection{The EXPLORE II Search}

The EXPLORE II search was carried out at the CFHT using the CFH12K camera
on 21/Dec/2001 to 05/Jan/ 2002, for a total of 16 nights.
(The actual assignment was 14 nights, which was shared over 16 nights
with another program.)
We obtain good coverage under mostly excellent conditions for 14 nights.
The CFH12K camera has a mosaic CCD detector of 12288$\times$8096 pixels
with a scale of 0.206$''$ per pixel, providing a field of view of $\sim 42'\times
28'$.
We monitored a field on the Galactic Plane at $l=207^\circ$ and $b=0.7^\circ$,
using an $I$ filter.
The fiducial integration time 
was 90 seconds for the nominal 0.8$''$ seeing.
With a readout time of 70 seconds, the sampling time is 2.7 minutes,
similar to EXPLORE I.
To avoid saturating the brighter stars,
the integration time was adjusted throughout the run from 55 to 120 seconds, 
depending on seeing, which ranged from 0.5$''$ to above 1$''$.
A total of $\sim2600$ science frames were obtained during the run.
The data were reduced and relative photometry light curves produced using
the same pipeline as that for EXPLORE I.
Because of the better seeing, longer integration time, and smaller 
pixel size, the average signal-to-noise (S/N) ratio is somewhat 
superior to that of the EXPLORE I data.
However, because of the limitation on the accessible RA range due to
the date of the observing run, the observable
Galactic Plane region has considerably lower star density than that
of the EXPLORE I search.
To maximize the star density, the field chosen also contains an open
star cluster, which occupies about two of the twelve chips of
the CCD.
Preliminary results from EXPLORE II are based on $\sim$9500 stars with rms
scatter of $<$1\%.
An example of the distribution of rms scatter (averaged over a whole
night) as a function of magnitude is shown in Figure 4.
Examples of light curves from the EXPLORE II search
are shown in Figure 5, and possible
planet candidates are discussed in Section 4.

\subsection{Spectroscopic Follow-up}

At this point, we have conducted a limited amount of spectroscopic 
follow-up of the planet candidates from EXPLORE I and II, using
the VLT 8.2m and Keck 10m telescopes, respectively.
We obtained 19 hrs of VLT Director's Discretionary time on September 2001
for radial velocity measurements of the EXPLORE I candidates.
We followed-up our three most promising candidates using UVES.
The observations were done in service mode, with data obtained 
over several nights, allowing for reasonable
phase coverages for the candidates.

For EXPLORE II, we were assigned 5 half-nights  on the Keck 
telescope with the HIRES spectrograph on February 2002, 
six weeks after the photometric survey.
Prior the the Keck run, we also obtained classification spectra
of our candidates on the APO ARC 3.5m, which
allowed us to further refine our radial velocity follow-up sample.
Due to poor weather, we obtained limited radial velocity data for two possible
planet candidates and two additional eclipsing stars at Keck.
Preliminary results of the RV follow-up observations are discussed
in conjunction with the
photometric light curves in the next section.

   \begin{figure}
   \begin{center}
   \begin{tabular}{c}
   \includegraphics[height=7cm]{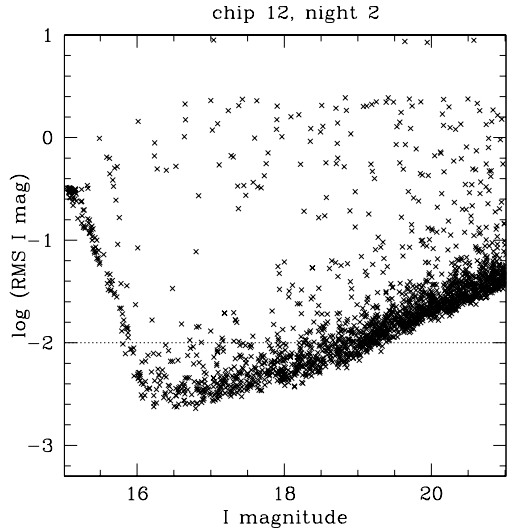}
   \end{tabular}
   \end{center}
   \caption[photrms]
   { \label{fig:photrms}
Typical photometric precision from the EXPLORE II search using the
CFH12k camera.
The rms in the photometry for each object is computed using measurements
from the whole night.
}
   \end{figure}

\section{Example Light Curves and Possible Planet Candidates} 

We present a number of interesting light curves from the two
EXPLORE searches, most with some velocity information.
These light curves also provide some indications of the quality of the
photometric data obtained by the EXPLORE project.
We show first three non-planet transiting/eclipsing systems, which 
provide interesting illustrations of the possible
contaminants discussed in Section 2.3.
We then present two of the four possible planet candidates we have
from our preliminary analysis of the two searches.
These are promising candidates with at least some velocity information, 
but all require additional data to 
confirm their status definitively.

\subsection{Non-Planet Transiting/Eclipsing Systems}
\noindent
(1) EXP1-c07s5157:~~~
This light curve from the EXPLORE I search shows two 3\%~eclipses
with a period of $\sim$3.2 days (Figure 3, left panel).
The high time resolution and precision of the light curve indicate
that this is likely a grazing binary with characteristic V-shape
eclipses.

\noindent
(2) EXP2-c04s5494:~~~
The left panel of Figure 5 shows a light curve from the EXPLORE II search
with a very prominent flat-bottomed eclipse of 3\%~depth.
The period, possibly 4.2 days, is not well determined, and
is based on a second potential eclipse partially observed during 
a night without full coverage.
The flat bottom means that the transiting body is 
completely superimposed on the primary star.
However, the long duration the eclipse 
suggests that this is a system with a small star eclipsing a
considerably larger star.
This star was not considered as a transiting planet candidate; however,
because of its brightness,
it was observed during the Keck spectroscopy run as a back-up object
during poor observing conditions.
Keck HIRES spectroscopy data show a single cross-correlation velocity peak which
shifts significantly, confirming that this is an eclipsing
binary system.

   \begin{figure}
   \begin{center}
   \begin{tabular}{c}
   \includegraphics[height=9cm]{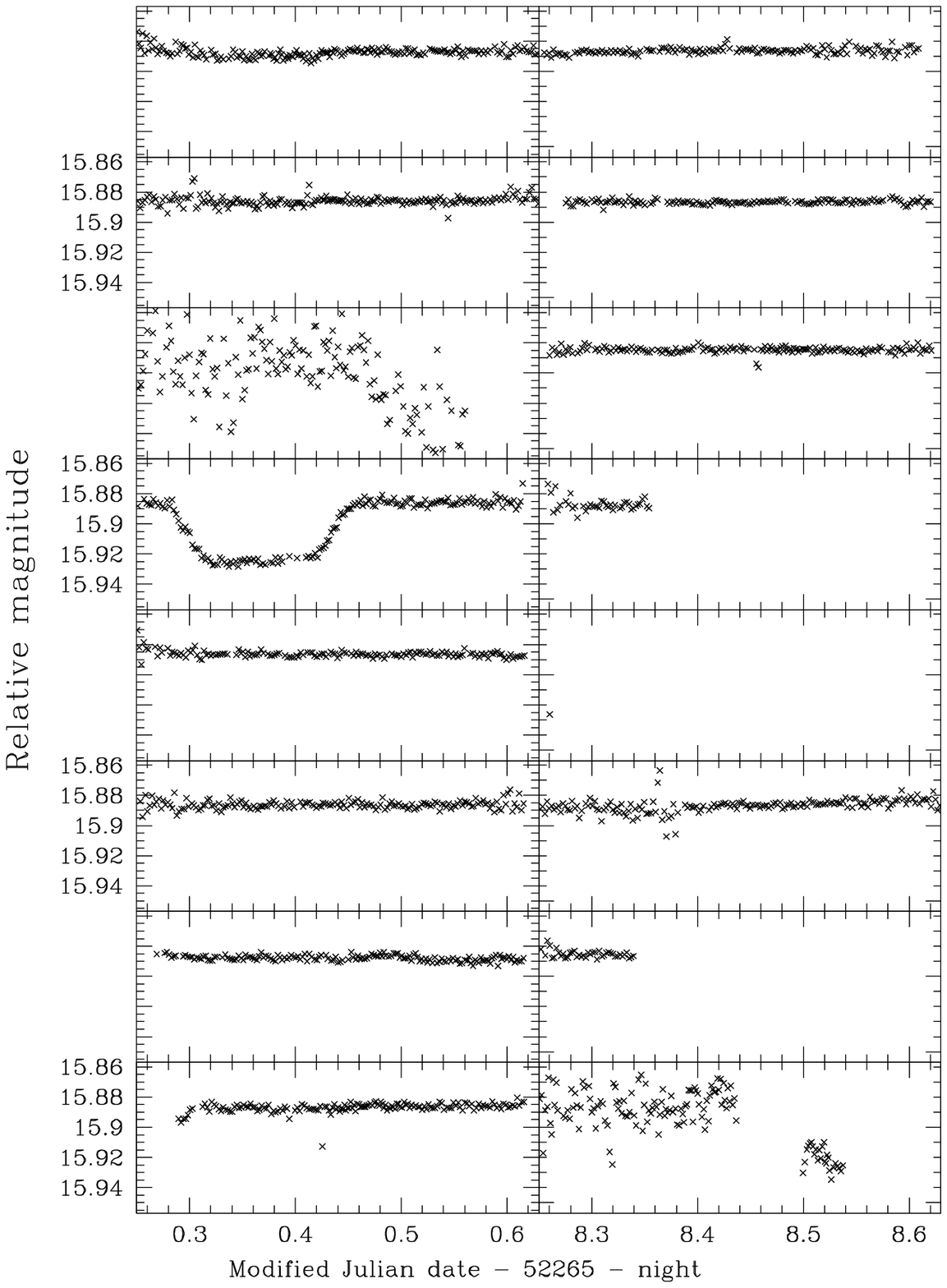}
   \includegraphics[height=9cm]{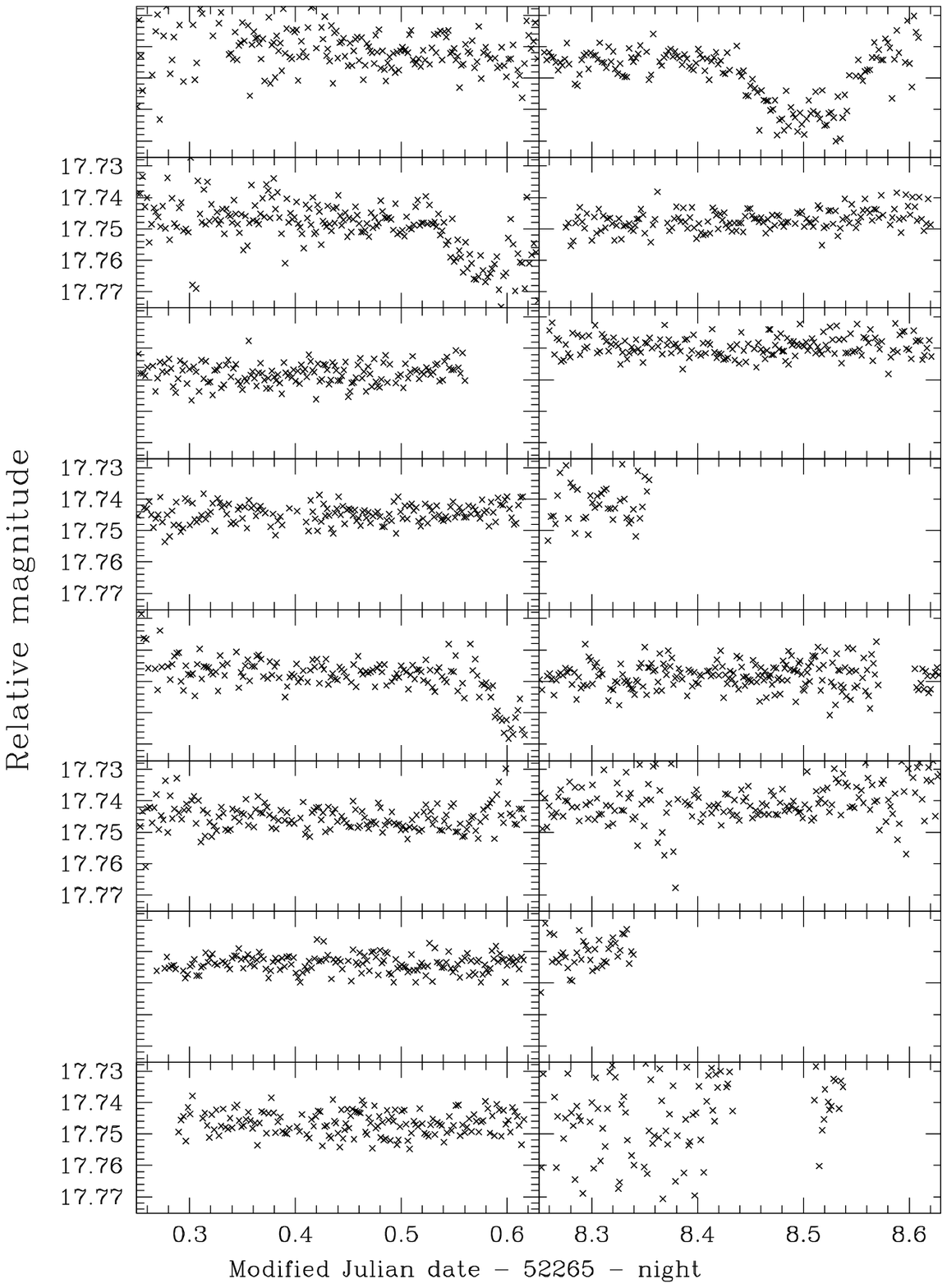}
   \end{tabular}
   \end{center}
   \caption[5494lc]
   { \label{fig:4809lc}
Examples of light curves from the EXPLORE II search carried-out
using the CFHT CFH12k camera.
The set up of the panel is the same as that of Figure 3.
The night numbers increase from bottom to top, with the
left column being nights 1 to 8, and the right column being
nights 9 to 16.
Left Panel: EXP2-c04s5494:
A star with a very prominent flat-bottomed eclipse of $\sim3$\% depth.
The long transit time indicates that it is likely a small star
crossing a large primary star.
Keck spectroscopy indicates significant radial velocity changes, 
supporting such a conclusion.
Right Panel: EXP2-c11s4809:
A possible planet candidate showing a flat-bottomed eclipse of 1.7\%
depth.
Currently, we have only very limited Keck spectroscopy data on this
object and do not have sufficient multi-phase radial velocity data for
the verification of the status of this candidate.
}
   \end{figure}

\noindent
(3) EXP1-c01s52805:~~~
This star from the EXPLORE I search shows two well observed flat-bottomed
eclipses with a depth of $\sim3$\%, and a period of 2.23 days
(Figure 3, right panel).
This was one of the three candidates for the VLT follow-up.
However, the relatively small ratio of the flat-bottom to ingress/egress
is a cause for concern.
Classification spectra indicate that this is an early K-star.
However, applying the uniqueness criterion to the light curve,
the best fit indicates a parent stellar radius of an F star.
Alternatively, this could be an eclipsing system of relatively late-type
stars blended with a brighter star.
VLT-UVES spectroscopy shows that there are in fact two cross-correlation
velocity peaks, a strong stationary one, and a second broad, very weak one consistent
with having a 2.23 day period and a $\sim60$ km s$^{-1}$ amplitude\cite{exp1},
indicating a three-star blended system.

\subsection{Possible Planet Candidates}

\noindent
(1) EXP2-c11s4809:~~~
This star from the EXPLORE II search
 shows a light curve with 
one full eclipse with a flat bottom and two partial eclipses
(Figure 5, right panel).
The eclipses have relatively short ingress/egress time, 
a period of 2.97 days and a depth of 1.7\%.
Based on the light curve, this is a promising candidate for RV follow-up.
We were only able to obtain two spectroscopic observations at Keck for this
possible planet candidate.
The data show only one radial velocity peak (and hence it is unlikely
a blend); however, both RV observations were obtained at almost the
same orbital phase, and hence we are not able to set a mass limit
for the transiting body.

\noindent
(2) EXP1-c07s18161:~~~
The light curve of this relatively faint star 
(Figure 6, left panel) from the EXPLORE I search shows two rather noisy
eclipses which are consistent with flat-bottomed transits with
a period of 3.8 days and a depth of $\sim2.5$\%.
This object was observed using the VLT-UVES spectrograph.
The preliminary radial velocity points
are shown in the right panel of Figure 6.
The spectroscopic data are consistent with showing no
radial velocity variations greater than $\pm200$ km/s, providing a 
preliminary mass
limit of $\sim$2.5 $M_J$, assuming that the eclipses observed in the
noisy light curves are real.
Additional photometric data are required to verify the
reality of the eclipses for a definitive confirmation of
this object as a transiting planet.

   \begin{figure}
   \begin{center}
   \begin{tabular}{c}
   \includegraphics[height=9cm]{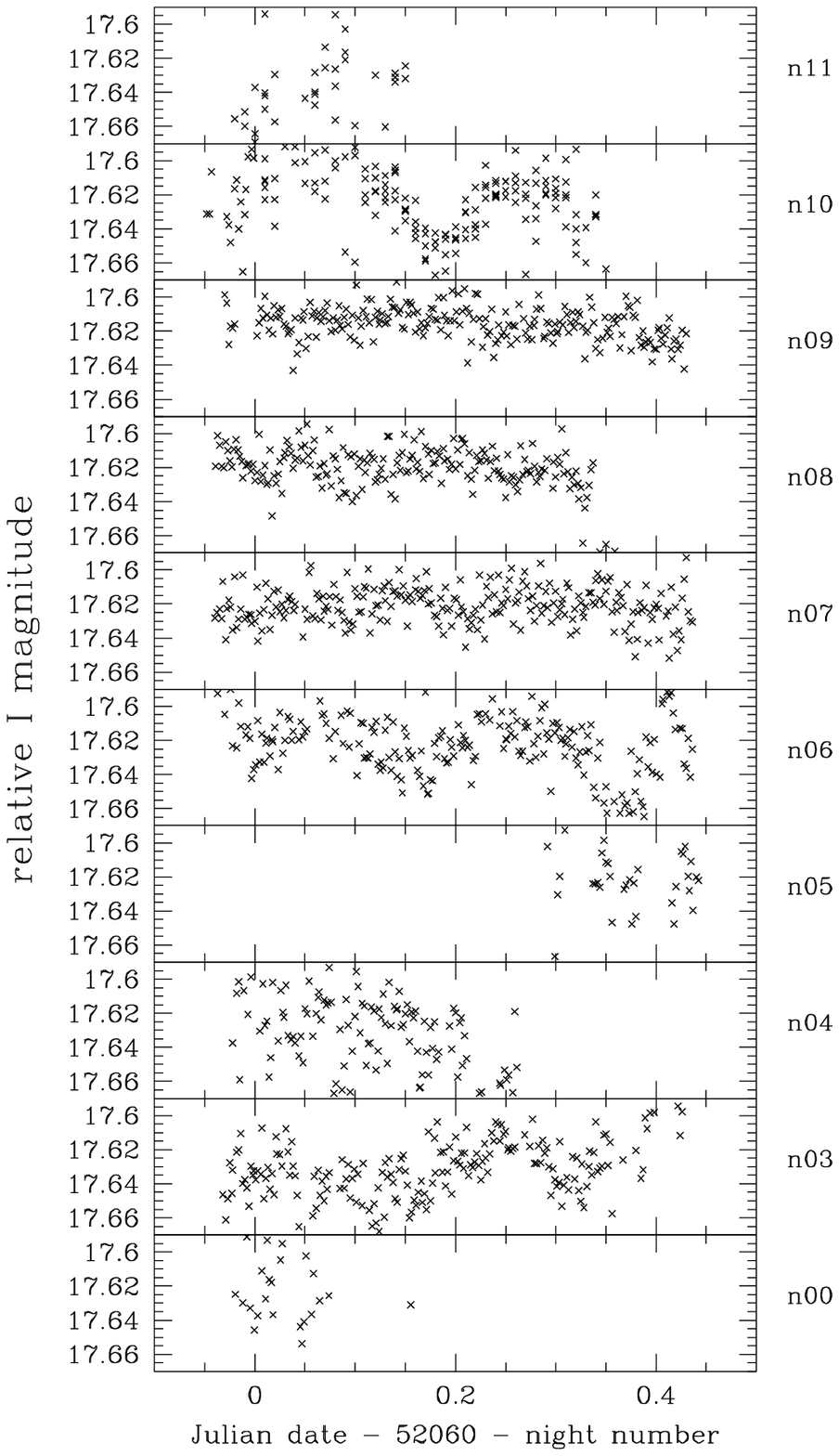}
   \includegraphics[height=7cm]{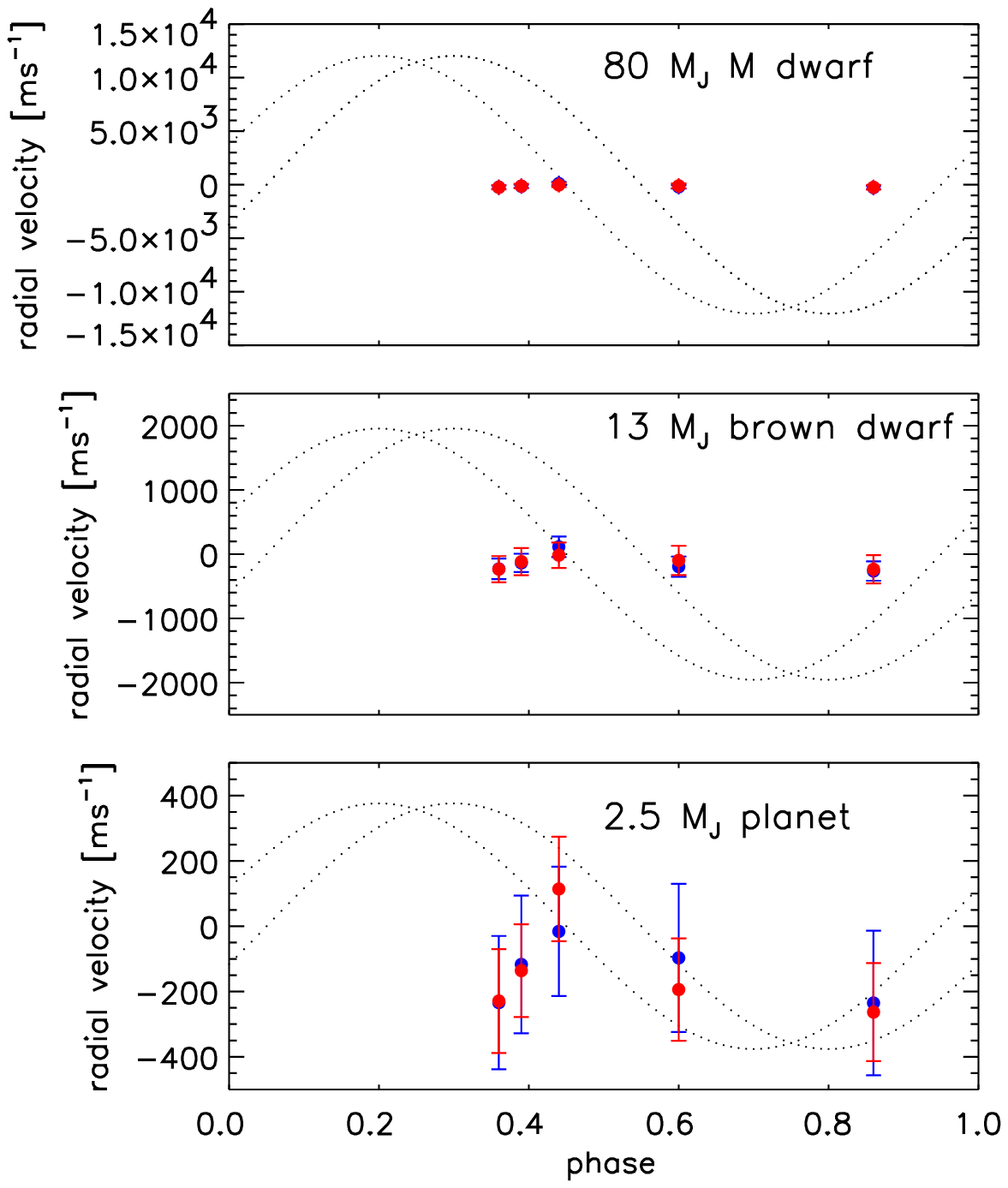}
   \end{tabular}
   \end{center}
   \caption[109lc]
   { \label{fig:109lc}
Left Panel:
The light curve of EXP1-c07s18161 from the EXPLORE I search,
a fainter star with a relatively low signal-to-noise ratio light
curve.
There are two possible flat-bottomed eclipses, though the one at
night 6 is very noisy.
Radial velocity data for this possible planet candidate
 were obtained using VLT-UVES.
The left panel shows the preliminary velocity data superimposed on expected
velocity models of an 80 $M_J$ M dwarf, a 13$M_J$ brown dwarf,
and a 2.5$M_J$ planet, with the uncertainty in the phase indicated.
The two data points for each phase are based on the red and blue 
parts of the spectrum.
}
   \end{figure}

\section{Summary and Future Prospects} 

Planet transit surveys have the promise of providing the next breakthrough
in extra-solar planet detection and characterization.
Transit searches can potentially yield a large number of close-in planets,
and such a sample will allow extra-solar planets to be characterized 
in much greater detail than is possible with non-transiting planets discovered by
RV techniques.
However, up to now, the transit search technique has not been successful in
producing new detections of planets.
We examined the various aspects in the design of transit searches, 
focusing on optimizing searches which  use telescopes with limited
time availability (e.g., 4m-class national facilities) and the
requirements for producing
high-yield samples for spectroscopic follow-up RV observations
using 8-10m class telescopes.

We showed that high-quality light curves with high precision
photometry and frequent
time sampling can provide sufficient information to winnow out most
of the contaminating objects which may mimic transiting planets.
We demonstrated that 8-10m class telescopes can provide useful
mass limits for these possible planet candidates even for relatively
faint stars of $I\sim18$ mag.
The advent of wide-field mosaic CCD cameras has made transit searches
very attractive.
We have conducted two searches using 4-m class telescopes as initial
surveys.  These two searches, with $\sim47,000$ preliminary 
light curves examined, have
produced four possible planet candidates which still require additional 
spectroscopic or photometry data for definitive confirmation.

New imaging capabilities at various observatories will further improve
the efficiency of transit searches in the near future.
One such example is the MegaCam at CFHT which will become operational in
2003.
With a 1 square degree field, small pixels, excellent image quality and
a very short readout time, it will be able to gather data for transit
searches at a much greater rate than currently possible.
For transit searches for fields at the Galactic Plane, the MegaCam can
routinely monitor as many as 120,000 stars simultaneously with 
photometric precision of better than 1\%.
Currently there are a large number of on-going transit surveys, and
large samples of transiting planets are expected to be discovered in
the near future.
The EXPLORE project is continuing with additional searches in 2002 and 2003
using NOAO 4m telescopes under their Survey Program.
As MegaCam becomes available, we also plan to propose to conduct additional
searches using CFHT.

\acknowledgments     
We would like to thank the staff at both CTIO and CFHT for their
very helpful assistance during the imaging runs, especially in accommodating the
very large data flows.
D.M.~is supported by FONDAP Center for Astrophysics 15010003.
S.S.~is supported by the W.M. Keck Foundation.
H.K.C.Y.'s research is supported by grants from NSERC of Canada
 and the University of Toronto.


\bibliography{expspie}   

\begin{thebibliography}{10}

\bibitem{mayor95}
M.~Mayor and D.~Queloz, ``A {J}upiter-mass companion to a solar-type star,''
  {\em Nature} {\bf 378}, p.~355, 2002.

\bibitem{lin96}
D.~N.~C. Lin, P.~Bodenheimer, and D.~C. Richardson, ``Orbital migration of the
  planetary companion of 51 {P}egasi to its present location.,'' {\em Nature}
  {\bf 380}, p.~606, 1996.

\bibitem{holman97}
M.~Holman, J.~Touma, and S.~Tremaine, ``Chaotic variations in the eccentricity
  of the planet orbiting 16 {CYG B},'' {\em Nature} {\bf 386}, p.~254, 1997.

\bibitem{murray98}
N.~Murray, B.~Hansen, M.~Holman, and S.~Tremaine, ``Migrating planets,'' {\em
  Science} {\bf 279}, p.~69, 1998.

\bibitem{rasio96}
F.~A. Rasio and E.~Ford, ``Dynamical instabilities and the formation of
  extrasolar planetary systems,'' {\em Science} {\bf 274}, p.~954, 1996.

\bibitem{seager00}
S.~Seager and D.~D. Sasselov, ``Theoretical transmission spectra during
  extrasolar giant planet transits,'' {\em ApJ} {\bf 537}, 2000.

\bibitem{char02}
D.~Charbonneau, T.~M. Brown, R.~W. Noyes, and R.~L. Gilliland, ``Detection of
  an extrasolar planet atmosphere,'' {\em ApJ} {\bf 568}, p.~377, 2002.

\bibitem{brown01}
T.~M. Brown, D.~Charbonneau, R.~L. Gilliland, R.~Noyes, and A.~Burrows,
  ``Hubble {S}pace {T}elescope time-series photometry of the transiting planet
  of {HD} 209458,'' {\em ApJ} {\bf 552}, p.~699, 2001.

\bibitem{vulcan}
W.~J. Borucki, D.~Caldwell, D.~G. Koch, L.~D. Webster, J.~M. Jenkin, Z.~Ninkov,
  and R.~Showen, ``The {V}ulcan photometer: A dedicated photometer for
  extrasolar planet searches,'' {\em PASP} {\bf 113}, p.~439, 2001.

\bibitem{stare}
T.~M. Brown and D.~Charbonneau, ``The {STARE} project; a transit search for hot
  {J}upiter,'' in {\em Planetary systems in the universe: observation,
  formation and evolution, {IAU} {S}ymp. 202},  A.~{Penny et al.}, ed., {\em
  ASP Conf. Ser.}, in press, 2002.

\bibitem{ogle}
A.~Udalski, K.~Zebrun, M.~Szymanski, M.~Kubiak, I.~Soszynski, O.~Szewczyk,
  L.~Wyrzykowski, and G.~Pietrzynski, ``The {O}ptical {G}ravitational {L}ensing
  {E}xperiment. search for planetary and low-luminosity object transits in the
  galactic disk. results of 2001,'' {\em Acta Astronomica} {\bf 52}, p.~115,
  2002.

\bibitem{char00}
D.~Charbonneau, T.~M. Brown, D.~W. Latham, and M.~Mayor, ``Detection of
  planetary transits across a sun-like star,'' {\em ApJ} {\bf 529}, p.~L45,
  2000.

\bibitem{henry00}
G.~W. Henry, G.~W. March, R.~P. Butler, and S.~S. Vogt, ``A transiting 51
  {P}eg-like planet,'' {\em ApJ} {\bf 529}, p.~L41, 2000.

\bibitem{exp1}
G.~Mall\'en-Ornelas, S.~Seager, H.~K.~C. Yee, D.~Minniti, M.~D. Gladders, G.~M.
  Mall\'en-Fullerton, and T.~M. Brown, ``The {EXPLORE} project {I}: a deep
  search for transiting extrasolar planets,'' {\em submitted to ApJ} {\bf
  astro-ph/0203218}, 2002.

\bibitem{unique02}
S.~Seager and G.~Mall\'en-Ornelas, ``On the unique solution of planet and star
  parameters from an extrasolar planet transit light curve,'' {\em submitted to
  ApJ} {\bf astro-ph/0206228}, 2002.

\bibitem{butler01}
R.~P. Butler, G.~W. Marcy, D.~A. Fischer, S.~Vogt, C.~G. Tinney, H.~R.~A.
  Jones, A.~J. Penny, and K.~Apps, ``Statistical properties of extrasolar
  planets,'' in {\em Planetary systems in the universe: observation, formation
  and evolution, {IAU} {S}ymp. 202},  A.~{Penny et al.}, ed., {\em ASP Conf.
  Ser.}, in press, 2002.

\bibitem{hubbard02}
W.~Hubbard, A.~Burrows, and J.~Lunine {\em ARAA} , in press, 2002.

\bibitem{blee02}
B.~Lee, H.~K.~C. Yee, G.~Mall\'en-Ornelas, and S.~Seager, ``Photometric
  simulations of transiting extrasolar planets for the {EXPLORE}
  observations,'' in {\em Scientific Frontiers in Research on Extrasolar
  Planets},  {\em ASP Conf. Ser.}, in press, 2002.

\end{thebibliography}
\bibliographystyle{spiebib}   

\end{document}